\documentclass[floatfix,showkeys,aps,prmat,reprint,superscriptaddress,longbibliography, twocolumn]{revtex4-1}

 \usepackage{epsfig}
 \usepackage{latexsym}
 \usepackage{epic}
 \usepackage{eepic}
 \usepackage{psfrag}
 \usepackage{rotating}
 \usepackage{multirow,booktabs}
 \usepackage{color}
 \usepackage{cprotect} 
 \usepackage{miller} 
 \usepackage{xspace} 
\usepackage[colorinlistoftodos,prependcaption,textsize=tiny]{todonotes}
\usepackage{graphicx}

  \usepackage[fleqn, leqno]{amsmath}
\usepackage{cleveref}  

\usepackage{amsfonts}  
  \usepackage{amssymb}   
 \usepackage{amsthm}
  \usepackage{stmaryrd} 
  \usepackage{import}

  \usepackage{array}

  \usepackage[normalem]{ulem}

\renewcommand\vec{\mathbf}

\newcommand{\MXfour}{X4 \xspace}
\newcommand{\ME}{E2 \xspace}

\newcommand{\ddk}{\ensuremath{{\mathrm d}{\mathbf k} }\xspace} 
 
\newcommand{\ddr}{\ensuremath{{\mathrm d}{\mathbf r} }\xspace} 
\newcommand{\rr}{\ensuremath{{\mathbf r} }\xspace}

\newcommand{\kk}{\ensuremath{{\mathbf k} }\xspace} 
\newcommand{\mm}{\ensuremath{{\mathbf m} }\xspace} 

\newcommand{\phii}[1][]{\ensuremath{\varphi}\xspace\ifx\relax#1\relax\else\ensuremath{\left(#1\right)}\xspace\fi}
\newcommand{\phih}[1][]{\ensuremath{\widehat{\varphi}}\xspace\ifx\relax#1\relax\else\ensuremath{\left(#1\right)}\xspace\fi}
 
\newcommand{\phiav}{\ensuremath{\bar{\phii}}\xspace} 

\newcommand{\qo}{\ensuremath{q_0}\xspace} 

\newcommand{\FF}[1][]{\ensuremath{\mathcal{F}}\xspace\ifx\relax#1\relax\else\ensuremath{\left[#1\right]}\xspace\fi}
\newcommand{\FFm}[1][]{\ensuremath{\mathcal{F_{\mathrm m}}}\xspace\ifx\relax#1\relax\else\ensuremath{\left[#1\right]}\xspace\fi}
\newcommand{\FFpfc}[1][]{\ensuremath{\mathcal{F_{\mathrm{PFC}}}}\xspace\ifx\relax#1\relax\else\ensuremath{\left[#1\right]}\xspace\fi}
\newcommand{\FFcoup}[1][]{\ensuremath{\mathcal{F_{\mathrm c}}}\xspace\ifx\relax#1\relax\else\ensuremath{\left[#1\right]}\xspace\fi}

\newcommand{\FFex}[1][]{\ensuremath{\mathcal{F_{\mathrm{exc}}}}\xspace\ifx\relax#1\relax\else\ensuremath{\left[#1\right]}\xspace\fi}

\newcommand{\ff}[1][]{\ensuremath{f}\ifx\relax#1\relax\else\ensuremath{\left(#1\right)}\fi}

\newcommand{\ffpfc}[1][]{\ensuremath{\ff_{\mathrm{PFC}}}\ifx\relax#1\relax\else\ensuremath{\left(#1\right)}\fi}
\newcommand{\ffid}[1][]{\ensuremath{\ff_{\mathrm{id}}}\ifx\relax#1\relax\else\ensuremath{\left(#1\right)}\fi}
\newcommand{\ffex}[1][]{\ensuremath{\ff_{\mathrm{ex}}}\ifx\relax#1\relax\else\ensuremath{\left(#1\right)}\fi}
\newcommand{\ffm}[1][]{\ensuremath{\ff_{\mathrm m}}\ifx\relax#1\relax\else\ensuremath{\left(#1\right)}\fi}
\newcommand{\ffcoup}[1][]{\ensuremath{\ff_{\mathrm c}}\ifx\relax#1\relax\else\ensuremath{\left(#1\right)}\fi}

\newcommand{\Ch}[1][]{\ensuremath{\widehat{C}}\xspace\ifx\relax#1\relax\else\ensuremath{\left[#1\right]}\xspace\fi}
\newcommand{\Ck}[1][]{\ensuremath{C_k}\xspace\ifx\relax#1\relax\else\ensuremath{\left[#1\right]}\xspace\fi}
\newcommand{\CPFC}[1][]{\ensuremath{C_\mathrm{PFC}}\xspace\ifx\relax#1\relax\else\ensuremath{\left[#1\right]}\xspace\fi}
\newcommand{\CkPFC}[1][]{\ensuremath{C_\mathrm{k,PFC}}\xspace\ifx\relax#1\relax\else\ensuremath{\left[#1\right]}\xspace\fi}
\newcommand{\NN}[1][]{\ensuremath{\mathcal{N}}\ifx\relax#1\relax\else\ensuremath{\left[{#1}\right]}\fi}
\newcommand{\NNj}[1][]{\ensuremath{\mathcal{N}_j}\ifx\relax#1\relax\else\ensuremath{\left[{#1}\right]}\fi}
\newcommand{\NNjp}[1][]{\ensuremath{\mathcal{N}_j^+}\ifx\relax#1\relax\else\ensuremath{\left[{#1}\right]}\fi}
\newcommand{\NNjm}[1][]{\ensuremath{\mathcal{N}_j^-}\ifx\relax#1\relax\else\ensuremath{\left[{#1}\right]}\fi}

\newcommand{\LL}[1][]{\ensuremath{\mathcal{L}}\ifx\relax#1\relax\else\ensuremath{#1}\fi}
\newcommand{\LLj}[1][]{\ensuremath{\mathcal{L}_j}\ifx\relax#1\relax\else\ensuremath{#1}\fi}
\newcommand{\GGj}[1][]{\ensuremath{\mathcal{G}_j}\ifx\relax#1\relax\else\ensuremath{#1}\fi}

\newcommand{\np}[1][]{\ensuremath{#1}\ifx\relax#1\relax\else\ensuremath{^{n+1}}\fi}
\newcommand{\n}[1][]{\ensuremath{#1}\ifx\relax#1\relax\else\ensuremath{^{n}}\fi}

\newcommand \be {\begin{eqnarray}}
\newcommand \ee {\end{eqnarray}}

\newcommand{\kkj}{\ensuremath{\vec{k}_j}\xspace}

\newcommand{\meS}[1][]{\ensuremath{\mathrm S}\ifx\relax#1\relax\else\ensuremath{_{#1}}\fi}

\newcommand{\MM}[1][]{\ensuremath{\mathcal{M}}\ifx\relax#1\relax\else\ensuremath{#1}\fi}
\newcommand{\MMj}[1][]{\ensuremath{\mathcal{M}_j}\ifx\relax#1\relax\else\ensuremath{#1}\fi}

\newcommand*\colvec[3][]{
    \begin{pmatrix}\ifx\relax#1\relax\else#1\\\fi#2\\#3\end{pmatrix}
}

\usepackage{geometry}

\input{MyBibaliasAcite.tex}
\begin{document}
\title{Magnetically enhanced thin film coarsening by a magnetic XPFC model allowing to decouple magnetic anisotropy and magnetostriction}
\author{Rainer Backofen*}
\affiliation{
  Institute of Scientific Computing, Technische Universit\"at Dresden, 01062 Dresden, Germany}
\email[]{rainer.backofen@tu-dresden.de}
\author{Axel Voigt}
\affiliation{
  Institute of Scientific Computing, Technische Universit\"at Dresden, 01062 Dresden, Germany}
\affiliation{Dresden Centre for Computational Materials Science (DCMS), TU Dresden, 01062 Dresden, Germany}

\begin{abstract}
External magnetic fields provide a macroscopic control mechanism to influence the microstructure of polycrystalline materials. We model the influence of strong magnetic fields on grain growth in thin films with a magnetic extended phase field crystal (XPFC) model. The magneto-structural effects are incorporated into the correlation function in reciprocal space. With this approach magnetic anisotropy, magnetostriction and mobility of grain boundary can be controlled and a variety of geometrical and topological properties consistent with experimental results can be determined. 
\end{abstract}

\keywords{Grain growth, magnetic driving force, grain selection}
\maketitle

\section{Introduction}

The properties of polycrystalline materials depend on the size- and shape-distributions of the crystallites they consist of. These distributions can be influenced by the conditions of preparation and subsequent processing. Understanding the grain coarsening process and the ability to predict the associated properties of polycrystalline materials are of high technological importance. They have been the subject of intensive experimental and theoretical research. 

A fairly simple approach to describe grain coarsening, due to Mullins  \acite{Mullins_JAP_1956}, assumes that it is mostly due to the motion of grain boundaries and triple junctions in a way that leads to the reduction of the excess free energy associated with the grain boundary network. While theories and simulations, based on this approach, predict the emergence of a time-invariant grain size distribution, which evolves in a self-similar way, and scaling laws of the average grain size $\langle A \rangle \sim t^\beta$, neither the distribution nor the scaling exponent $\beta = 1$  \acite{Mullins_AM_1998,BKS06} reproduce experimental results. In  \acite{Barmaketal_PMS_2013} it has been shown that geometric and topological characteristics of the grain structure are universal and independent of many experimental conditions, and the scaling exponent is typically smaller or scaling breaks down as coarsening stagnates. Various additional effects such as grain rotation, elasticity, and dislocation dynamics have been found to influence grain growth and models on a molecular scale, which consider these effects, reproduce the experimental results  \acite{BBV14,LaBoissoniereetal_M_2019}. The considered phase field crystal (PFC) model, introduced by Elder et al.  \acite{Elder2002,Elder2004}, contains a much richer physics than the previous models proposed by Mullins  \acite{Mullins_JAP_1956}. 

Magnetic fields offer a macroscopic control mechanism to influence grain growth  \acite{Guillonetal_MT_2018,Riv13,WTE06}. While various simulation attempts to predict the influence of magnetic fields on grain growth exist based on Mullins-type models \acite{BMM07,Koyama_STAM_2008,Leietal_EPL_2009,Allen_JEMTT_2016,Goinsetal_CMS_2018,ReZaeietal_CMS_2021}, their predictive power is questionable in view of the above discussion. Within a series of papers also the PFC model was extended to capture the fundamental physics of magnetocrystalline interactions  \acite{FPK13,SSP15,FMG18,BEV19,BV20,BSV22a,BSV22b}. Magneto-structural interactions are incorporated phenomenologically building on symmetry arguments. These models allow to simulate the effect of external magnetic fields on grain growth and were able to reproduce, at least qualitatively, the main results of experimental studies on thin polycrystalline structures \acite{MB10,BV20}. It was found that grain growth kinetics depend on the magnetic field direction. Grains with an energetically preferred orientation grow faster and their volume fraction becomes larger compared with grains with a disfavored orientation. However, in contrast to experimental results the simulations also predict an increased elongation of the grains perpendicular to the magnetic field. All these effects on the size- and shape-distributions of the crystallites and the scaling properties result from a complex interplay of magnetostriction, magnetic anisotropy and anisotropy of the grain boundary mobility with the rich physics already present without magnetic coupling. These couplings are partly explained in  \acite{BEV19,BV20}. However, the additional properties are not independent in the considered magnetic PFC model, which makes it challenging to control them or adapt them to specific materials. 

We propose a different coupling method that introduces magnetic anisotropy without magnetostriction and anisotropic grain boundary mobility. This is crucial to control the physical properties independently. In order to introduce this novel approach we briefly review the derivation of the PFC model from classical density functional theory (cDFT). Instead of approximating the correlation function in real space, as in the classical PFC model  \acite{Elder2002,Elder2004}, and adding a magnetic coupling term  \acite{FPK13,SSP15,BEV19,BV20}, we consider the XPFC model  \acite{GPR10} and directly modify the correlation function in Fourier space to account for the effect of magnetic anisotropy. We demonstrate that this novel approach reproduces the main magnetic properties of the previous approach with respect to the grain size distribution and the scaling behaviour. However, it differs with respect to the shape distribution. With the new approach, grain elongation can be controlled during coarsening. It therefore provides a feasible way to simulate quantitative grain growth taking magnetocrystalline interactions into account.

\section{Model}

\subsection{From cDFT to (X)PFC}
\begin{figure}[htb]
  \begin{tabular}{c}
    \includegraphics[width=0.45 \textwidth]{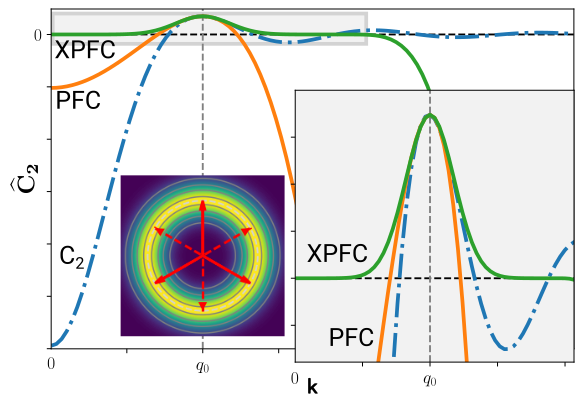}
  \end{tabular}
 \caption{PFC and XPFC approximation to the correlation function $\Ch_2$ for a triangular lattice. In (X)PFC the height, position and curvature of the first peak of $\Ch_2$ in reciprocal space is fitted. PFC uses a second order polynomial in $\kk^2$ and XPFC a Gaussian. In reciprocal space the maxima of $\Ch_2$ defines one circle. The red and the dashed red arrows define the basis modes describing the lattice and their negatives.}
  \label{fig:DFT2PFC}
\end{figure}

The PFC model can be derived from classical density functional theory (cDFT) \cite{EPG07,TBL09,Archeretal_PRE_2019,Pis13}. The free energy depends on the number density,
$\rho(\rr)$ and can be approximated as 
\begin{align}
  \nonumber
  \FF_{\rm cDFT}&[\rho(\rr)] = \int \rho(\rr) \ln(\rho(\rr)) \ddr \\
                &- \frac{1}{2} \int \rho(\rr) C_2(|\rr-\rr'|) \rho(\rr') \ddr' \ddr
                  \label{eq:FCDFT}
\end{align}
with $C_2$ the two point correlation function. The first part is the entropic energy and the second part the excess free energy. Within PFC models \cite{EPG07} a rescaled density deviation from a reference state is considered $\phii$, the first part is approximated by a polynomial and the second part expanded in reciprocal space. The free energy reads
\begin{align}
  \nonumber
  \FF_{\rm PFC}&[\phii[\rr] ] = \int \frac{u}{4}\phii[\rr]^4+\frac{t}{3} \phii[\rr]^3 \ddr \\
  &-\frac{1}{2} \int \phih[\kk ]  \Ch_{2}(|\kk|) \phih(\kk)   \ddk
\label{eq:FPFC0}
\end{align}
with $u$ and $v$ parameters to control, together with the mean density $\phiav$, the physical properties of the model and $\phih$ and $\Ch_2$ the Fourier representations \cite{OSP13}. To obtain the original PFC model of Elder et al. \cite{Elder2002,Elder2004} $\Ch_2$ is
approximated as a 4th order polynomial in $\kk$:
\begin{align}
  \Ch_{\rm PFC}(|\kk|)= & -(\epsilon+(\qo^2-\kk^2)^2),
                  \label{eq:CPFC}
\end{align}
with parameters $\epsilon$ and $q_0$, see Figure \ref{fig:DFT2PFC}. This approach leads to a local free energy
\begin{align}
  \nonumber
  \!\!\!\!\!\FFpfc&[\phii[\rr] ] = \int \frac{u}{4}\phii[\rr]^4+\frac{t}{3} \phii[\rr]^3 \ddr \\ 
  &+\frac{1}{2} \int \phii[\rr] (\epsilon+(q_0^2+\nabla^2)^2) \phii[\rr] \ddr
\label{eq:FPFC}
\end{align}
which is the basis for various real space methods to solve the PFC model, see e.g. \acite{Backofenetal_PML_2007,Wiseetal_SIAMJNA_2008,Dongetal_CMA_2018}.

A different approach is to model $\Ch_2$ as a set of
Gaussians directly in reciprocal space. This extended PFC (XPFC) model \cite{GPR10} no longer leads to a local representation in reals space but adds flexibility in the approximation of the correlation function. For every peak in $\Ch_2$ a Gaussian is defined
\begin{align}
  G_j(\kk)=g_j e^{-\frac{(q_j-|\kk|)^2}{\sigma_j}}
  \label{eq:GXPFC}
\end{align}
with parameters $g_j$, $q_j$ and $\sigma_j$, defining the position, the height and the width of the peak, respectively. $\Ch_2$ is approximated as 
\begin{align}
  \nonumber
  \Ch_{\rm XPFC} =\max(&G_0(\kk),G_1(\kk), \dots, G_n(\kk))  \\
  &-||\kk|-|\kk_d||^\alpha,
\label{eq:CXPFC}
\end{align}
with the last term introduced to suppress density fluctuations for large \kk with $|\kk|>|\kk_d|$. While this approach is not suited for real space methods it still allows for efficient solutions in Fourier space, see e.g. \cite{EW13,BBV14,Pis13,CW08}. Figure \ref{fig:DFT2PFC} shows a comparison of $\Ch_2$, $\Ch_{\rm PFC}$ and $\Ch_{\rm XPFC}$ for $n=0$, representing a triangular lattice, for which the basic length scale is preserved by approximating the first peak of the correlation function in reciprocal space. The red arrows correspond to the basic modes describing the lattice, they read
\begin{align}
\frac{\kkj}{q_0} \in
    \left\{\begin{bmatrix} 0 \\ 1 \end{bmatrix},
    \frac{1}{2} \begin{bmatrix} -\sqrt{3} \\ -1 \end{bmatrix},
    \frac{1}{2} \begin{bmatrix} \sqrt{3} \\ -1 \end{bmatrix}
     \right\}
\label{eq:kTria}
\end{align}  
and the dashed red arrows correspond to negative $\kkj$'s.

\begin{figure}[htb]
  \includegraphics[width=0.468 \textwidth]{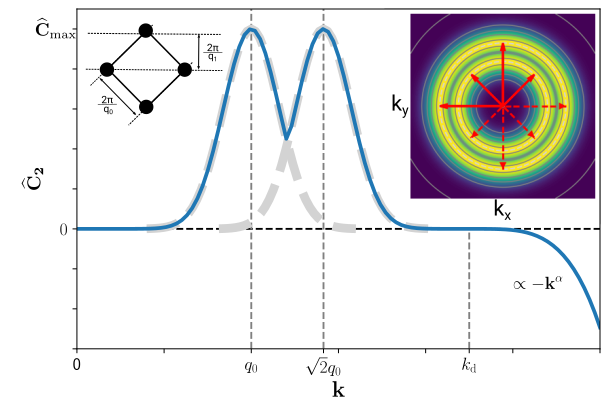}
  \caption{XPFC approximation for a square crystal. The considered length scales are $q_0$ and $\sqrt{2} q_0$, corresponding to the first two most basic crystallographic planes \cite{GPR10}. In reciprocal space the maxima of $\Ch_2$ define two circles. The red and the dashed red arrows define the basis modes describing the lattice and their negatives.}
  \label{fig:C2XPFC}
\end{figure}

In the following we will consider a square lattice. This can only be described by including more length scales or introducing nonlinear terms \cite{WPV10,MEH13,BSV21}. Introducing additional length scales leads to higher order terms in the PFC model \cite{MEH13} and thus increases the complexity. Within XPFC models this can be realized by considering two Gaussians ($n=1$ in eq. \eqref{eq:CXPFC}), see Figure \ref{fig:C2XPFC}.

The basic modes read
\begin{align}
\frac{\kkj}{\sqrt{2} q_0} \in
    \left\{ \frac{1}{2}  \begin{bmatrix} 1 \\ 1 \end{bmatrix},
    \frac{1}{2}  \begin{bmatrix} -1 \\ 1 \end{bmatrix},
    \begin{bmatrix} 0 \\ 1 \end{bmatrix},
     \begin{bmatrix} -1 \\ 0 \end{bmatrix}  \right\},
\label{eq:kSquare}
\end{align}  
where the first two correspond to the first peak and the third and forth vector correspond to the second peak, respectively the inner and outer circle in the correlation function, see Figure \ref{fig:C2XPFC}.


\subsection{Magnetic interaction}
\begin{figure}[htb]
  \begin{tabular}{cc}
    \includegraphics[width= 0.225 \textwidth]{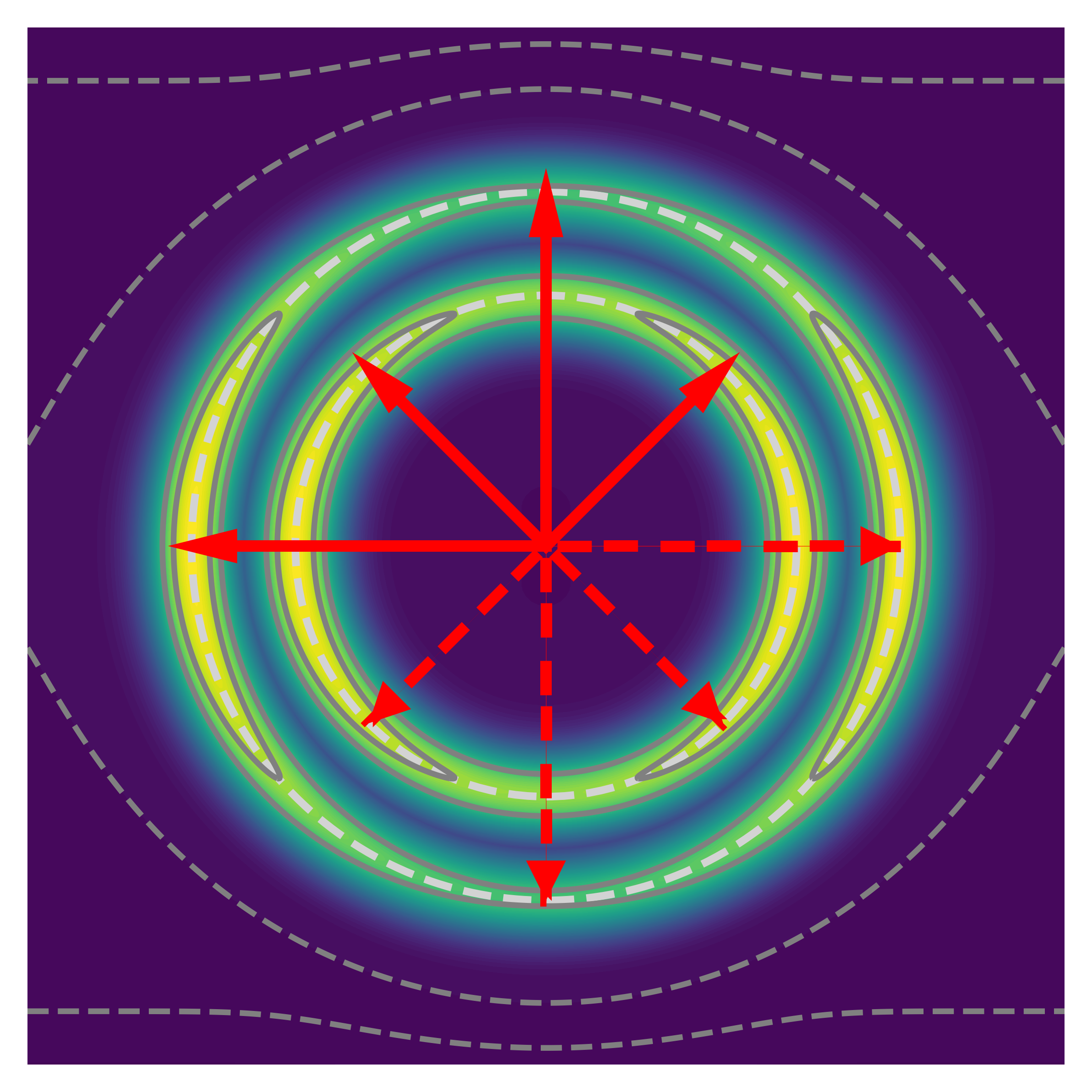} &                    \includegraphics[width= 0.225 \textwidth]{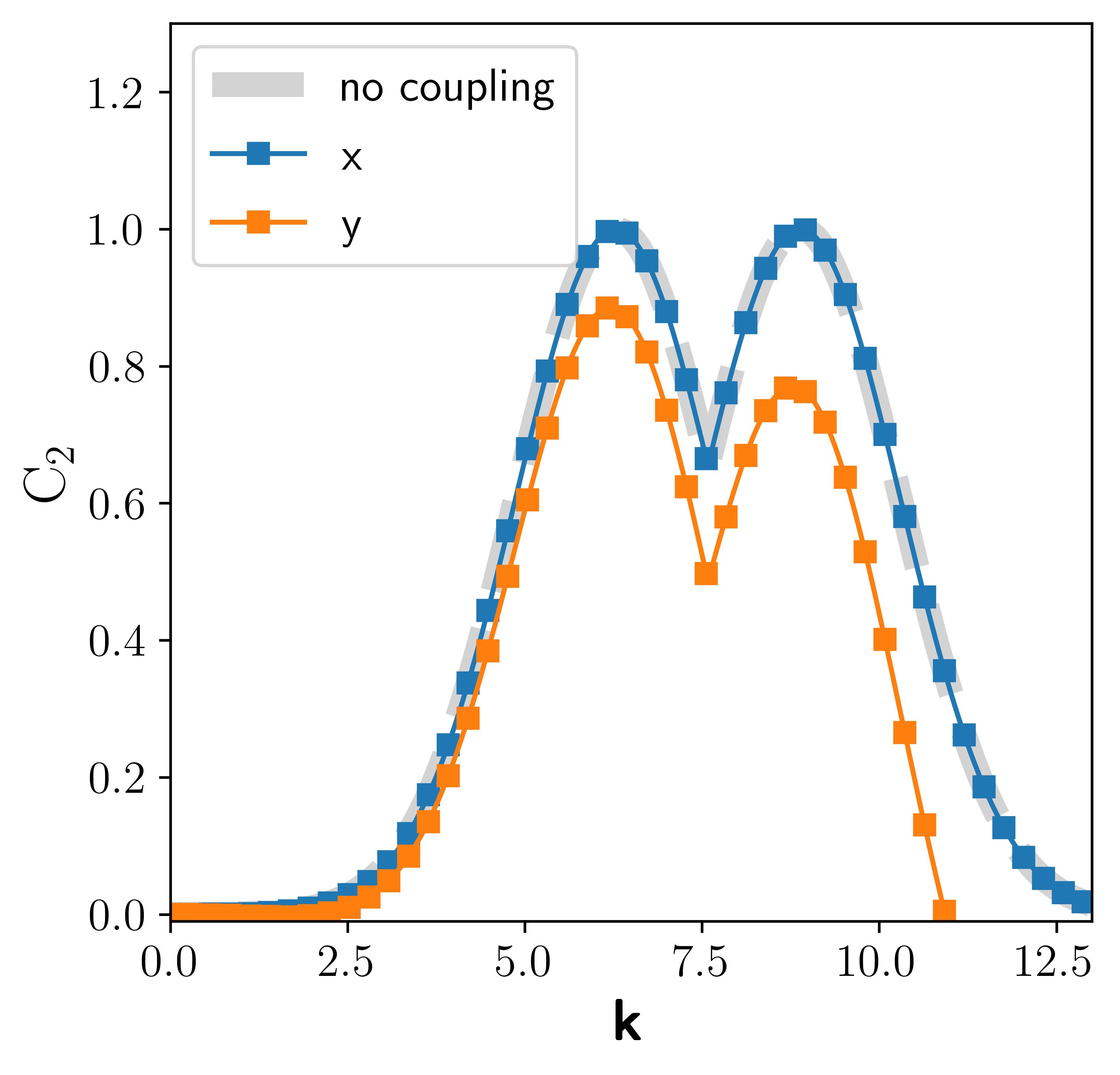} 
  \end{tabular}
  \caption{Magnetic coupling in XPFC (E2), as proposed by Backofen et. al. \cite{BSV22a}. (left): $C_2$ in reciprocal space. The red arrows are the \kkj's defining the square crystal. Dashed arrows correspond to negative \kkj's. The magnetic coupling can be viewed as an anisotropic correlation function in reciprocal space. The height and the position of the maxima of $\Ch_2$ is changed by the magnetic coupling. (right): Corresponding plot for $k_x$ and $k_y$. The plot considers $\mm$ in \hkl[11] direction pointing upwards.}
  \label{fig:C2E2}
\end{figure}

In the limit of strong magnetic fields the local magnetization is perfectly aligned with the external field and assumed to be constant in the crystal. 
Within this limit magnetocrystalline interactions can be considered by an additional energy contribution ${\cal{F}}_\mm[\phii(\rr), \mm]$. Different functional forms have been proposed. Seymour et al. \cite{SSP15} propose a formal expansion in $(\mm \cdot \nabla \phii)^{2j}$ and Backofen et al. \cite{BSV22b} propose an expansion in $\phii (\mm \cdot)^{2j} \phii$. Both expansions are equivalent in lowest order ($j = 1$), which is sufficient to describe magnetic anisotropy and magnetostriction in a 2D square lattice \cite{BEV19,BV20}. The ansatz in \cite{BSV22b} is computationally preferable and has only quadratic terms in \phii and thus can be interpreted the same way as the correlation function in reciprocal space. In lowest order the magnetically modified correlation function reads
\begin{align}
  \Ch_{\rm E2}(\kk)=\Ch_{\rm XPFC}(\kk) - a_2 (\mm \cdot \kk)^2 
        \label{eq:MPFC}
\end{align}
where $a_2$ defines the strength of magnetic coupling and \mm is rescaled such that $|\mm|=1$. The magnetic coupling introduces terms dependent on the direction of magnetization and, thus, breaks the rotational symmetry of the free energy. The energy of a crystal becomes dependent on its orientation w.r.t. \mm. It introduces a 2-fold symmetry in reciprocals space. We thus call this model expansion model with 2-fold symmetry (E2), see Figure \ref{fig:C2E2}.

\begin{figure}[htb]
  \begin{tabular}{cc}
    \includegraphics[width= 0.225 \textwidth]{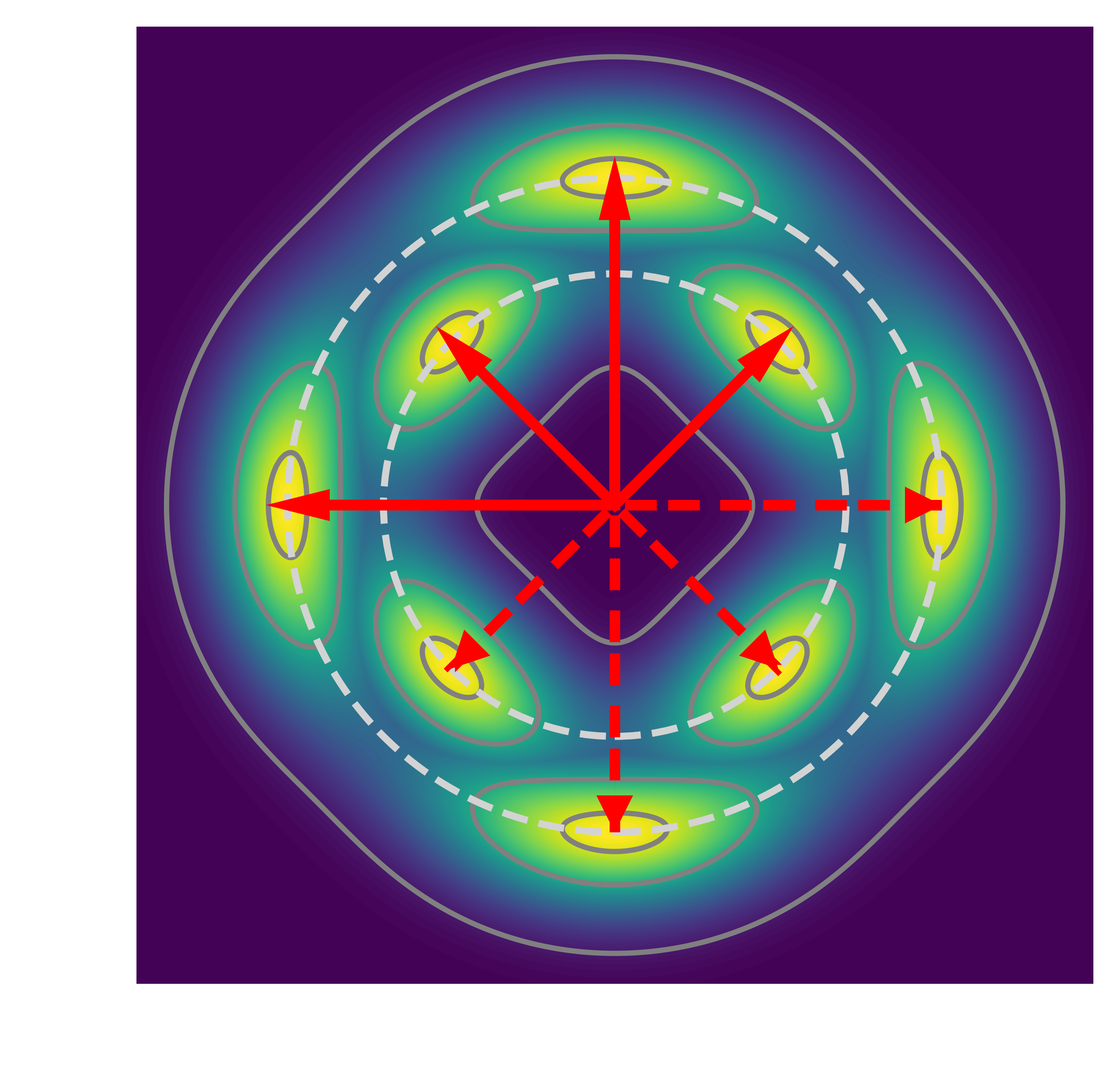} &                    \includegraphics[width= 0.225 \textwidth]{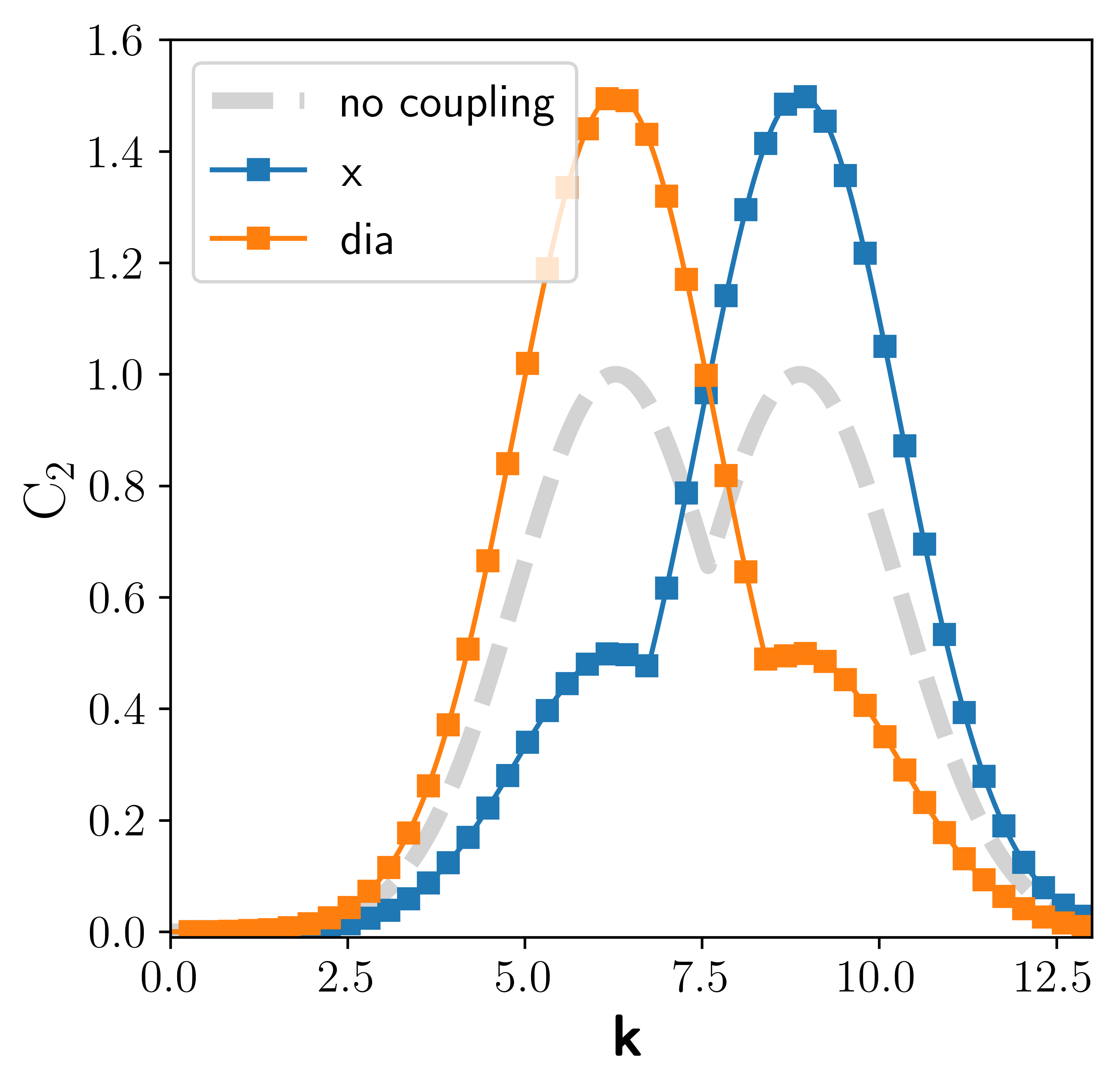}
  \end{tabular}
  \caption{Magnetic coupling in XPFC (X4). (left): $C_2$ in reciprocal space. The red arrows are the \kkj's defining the square crystal. Dashed arrows correspond to negative \kkj's. The magnetic coupling can be viewed as an anisotropic correlation function in reciprocal space. The height and the position of the maxima of $\Ch_2$ is changed by the magnetic coupling. (right): Corresponding plot for $k_x$ and in diagonal direction. The plot considers $\mm$ in \hkl[11] direction pointing upwards.}
  \label{fig:C2X4}
\end{figure}

Model E2 follows the idea of the original PFC model, expanding the energy in the lowest necessary order of the physical fields and their derivatives. But we can also introduce the magnetic coupling with the philosophy of the XPFC model, by directly defining the correlation function in reciprocal space. In order to break the symmetry the height of the Gaussians are varied dependent on the magnetization \mm. We consider
\begin{align}
  G_{{\rm mag},j}(\kk)=g_{{\rm mag},j}(\kk) e^{-\frac{(q_j-|\kk|)^2}{\sigma_j}}
  \label{eq:GMXPFC}
\end{align}
with
\begin{align}
 g_{{\rm mag},j}(\kk)= g_j+dg_j \cos(2 \theta)+g_{\rm s} 
\label{eq:gMXPFC}
\end{align}
where $g_j$ as in eq. \eqref{eq:GXPFC}, $dg_j$ is the variation of the height, $\theta$ is the angle between \mm and \kk and  $g_{\rm s}$ is a shift of the mean height of the maxima due to \mm. The magnetically modified correlation function reads
\begin{align}
  \nonumber
  \Ch_{\rm X4} =\max(&G_{{\rm mag},0}(\kk),G_{{\rm mag},1}(\kk))  \\
  &-||\kk|-|\kk_d||^\alpha
\label{eq:CMXPFC}
\end{align}
This model also breaks the rotational symmetry and has a 4-fold symmetry in reciprocal space. As this approach follows the XPFC  philosophy and has a 4-fold symmetry we call it (X4), see Figure~\ref{fig:C2X4}.

\subsection{Basic magnetic properties}

\begin{figure}[htb]
 \includegraphics[width= 0.45 \textwidth]{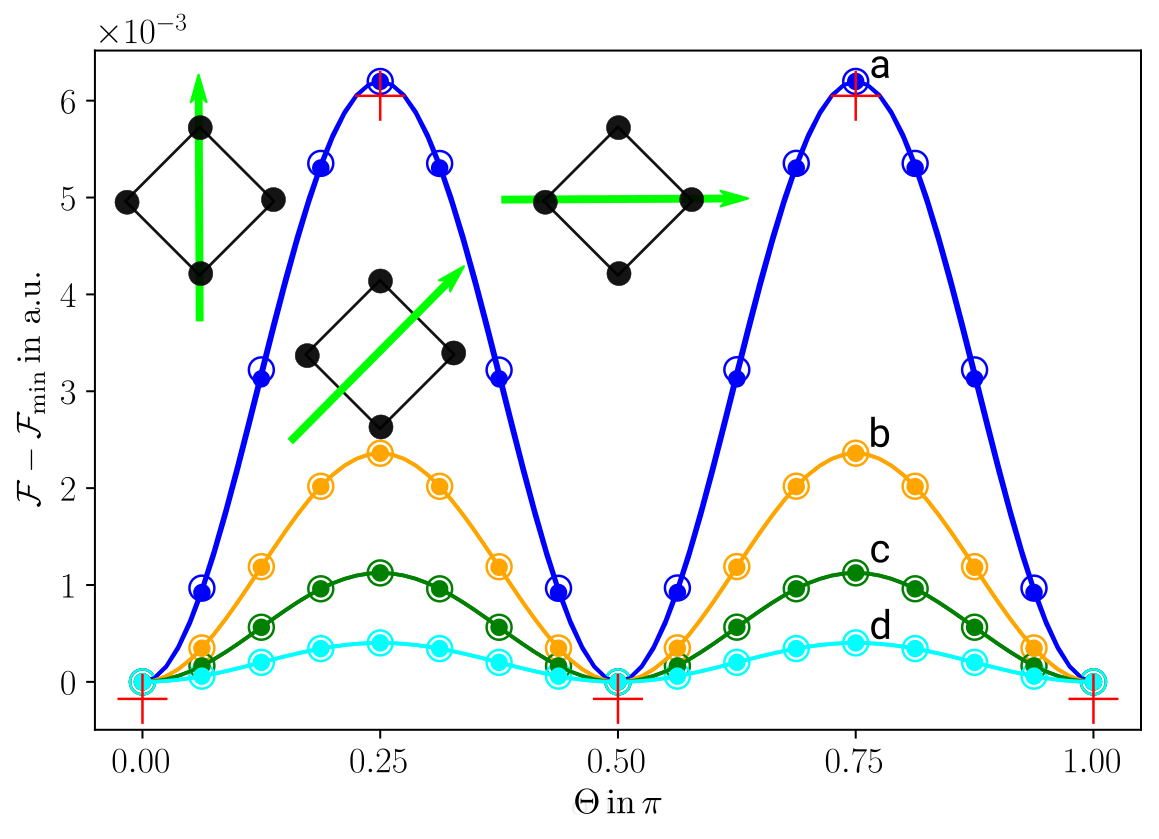}
  \caption{Magnetic anisotropy in E2 and X4 for four sets of parameters, see Table~\ref{tab:mpar}. The parameters in X4 (open symbols) are chosen, such that the magnetic anisotropy is the same as in E2 (closed symbols). Elastic relaxation has only a small effect on the magnetic anisotropy for E2 (red cross) and there is no magnetostriction for X4.}
  \label{fig:MAniso}
\end{figure}

  In order to analyze the magnetic properties of the models E2 and X4 a single crystal with constant magnetization is considered. The \kkj's are chosen to represent the crystal without magnetization. 
  
  Firstly, the simulation domain is fixed and no elastic relaxation of the crystal is possible. Then the energy is minimized for different directions of \mm, see Figure~\ref{fig:MAniso}. The model parameters for magnetic coupling are chosen such that both models have the same magnetic anisotropy, see Table~\ref{tab:mpar}. The free energy is minimal for \mm in \hkl<11>-directions and maximal for \mm in \hkl<10>-directions. Thus the energy in both models show the four fold symmetry of the square lattice. The magnitude of the magnetic anisotropy is controlled by changing $a_2$ in E2 and the height variation, $dg_i$, in X4.

\begin{table}[htb]
    \centering
    \begin{tabular}{c|c|ccc}
         & E2 & \multicolumn{3}{c}{X4} \\ \hline
        & $a_2$ & $dg_0$ & $dg_1$  & $g_s$ \\ \hline
    a     & -0.0021 & -0.002688& 0.002688 & -0.03262 \\
    b     & -0.0015 & -0.000928& 0.000928 & -0.01787 \\ 
    c    & -0.0012 & -0.000426 & 0.000426 &-0.01129 \\
    d     & -0.0009 & -0.000148 & 0.000148 &-0.00680 
    \end{tabular}
    \caption{Model parameters which lead to the same magnetic anisotropy in models E2 and X4, see Figure~\ref{fig:MAniso}. The square symmetry is achieved by: $(\phiav,u,t)=(0.05,-\frac{1}{2},\frac{1}{3})$, $(q_0,q_1)=(2 \pi, \sqrt{2} q_0)$ and $(\sigma_0,\sigma_1)=(g_0,g_1)=(1,1)$.}
    \label{tab:mpar}
  \end{table}

  Secondly, the crystal was allowed to relax elastically for hard and easy direction of \mm. There is no deformation of the crystal for X4. All the \kkj are on local extrema of $\Ch_2$, thus no infinitesimal change of \kkj change the magnetic excess free energy. This is different for E2. The excess free energy can be minimized by expanding the crystal in \mm direction \cite{BEV19}. Even though the magnetostriction is quite large, the influence on the magnetic anisotropy is low, see Figure-\ref{fig:MAniso}.     

\begin{figure}[htb]
 \includegraphics[width= 0.45 \textwidth]{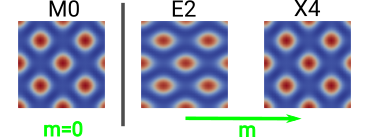}
  \caption{Influence of magnetic coupling on density distribution. Density peaks are elongated along magnetization direction in E2.}
  \label{fig:peaks}
\end{figure}

Apart from magnetostriction, there are other differences between the models. Figure~\ref{fig:peaks} shows the rescaled density field $\phii$ with and without magnetic interaction. The density peaks form a square crystal with \hkl<11> directions up and sideways. Due to the crystal's symmetry, density peaks are square-like rather than circular in the absence of magnetization. Magnetization in the crystal is oriented sideways in the \hkl[11] direction. In model E2, the density distribution is disturbed, and the density peaks are elongated along the magnetization direction. As a result, the symmetry of the density distribution is reduced. This density distribution reflects the 2-fold symmetry of the magnetic correlation function, see Figure~\ref{fig:C2E2}. However, as discussed above, the anisotropy of magnetic fields remains 4-fold symmetric. In model \MXfour, the magnetic correlation is particularly constructed in such a way that it reflects the 4-fold symmetry of the square lattice. Hence, the density distribution remains 4-fold symmetric and is not influenced by the direction of magnetization.

We now summarize the properties of the two models E2 and X4. Both models assume that the magnetization modifies the correlation function $\Ch_2$. The magnetic anisotropy is introduced by breaking the rotational symmetry of the correlation function in reciprocal space. The first model, E2, leads to a 2-fold correlation function in reciprocal space. This model leads to magnetic anisotropy, magnetostriction and deformation of the density distribution due to magnetization. The second model, X4, leads to a 4-fold correlation function in reciprocal space. Here magnetic anisotropy can be modeled without magnetostriction. Additionally the density distribution stays 4-fold symmetric. 

In the following we examine how the differences in these models influence grain growth and coarsening behavior.  

\section{Magnetically induced grain growth}

In this section, we apply the two models, E2 and X4, to study grain growth. We first consider two idealized situations, a magnetically driven planar tilt grain boundaries and the growth of a single circular grain. These situations highlight differences between the models. The second part considers coarsening of thin films under the influence of an external magnetic field. 

\subsection{Idealized grain growth}
Magnetic fields lead to additional driving forces on grain boundaries caused by the magnetic anisotropy present in the material \cite{AG89,TC94}. Using a Mullins-type model this reads:     
  \begin{align}
 \label{eq:vel}
    v_{12} = -M_\kappa \gamma \kappa - M_f \Delta f_{12} ,
\end{align}
where $v_{12}$ is the normal velocity of the grain boundary, $M_\kappa$ and $M_f$ are mobilities, $\gamma$ is the energy and $\kappa$ the mean curvature of the grain boundary.
The first term minimizes the energy by means of mean curvature flow. That is, it reduces the grain boundary. Magnetic anisotropy leads to a difference in energy density in the adjacent grains, which is expressed by $\Delta f_{12}$. As a result, a driving force is introduced to push the grain boundary towards the grain with the larger energy density.
\begin{figure}[htb]
    \includegraphics*[width= 0.45 \textwidth]{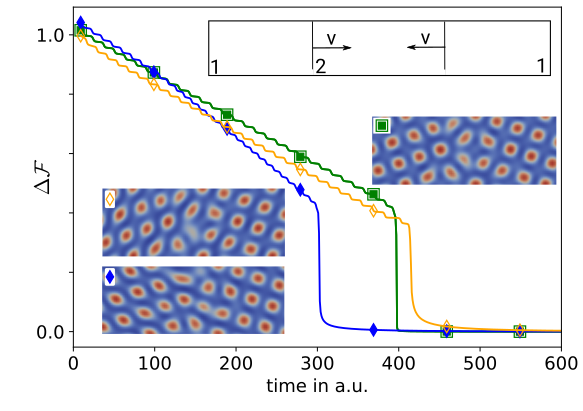}
    \caption{ \label{fig:gbVel} Magnetically driven planar grain boundary. Free energy decrease for two different setups with $\Delta \FF = (\FF-\FF_{\rm final})/\FF_{\rm initial}$. The initial and final free energy is measured for model \MXfour.  
The magnetization is in direction of an easy direction of the left grain. The magnetization is more parallel to the grain boundary for open symbols and more perpendicular for closed symbols. Both setups induce the same magnetic driving force. The squares correspond to model X4 and the diamonds to model E2. The configuration of the grain boundaries are shown in the inlays.  }
\end{figure}

As a first step, we consider a planar grain boundary driven by magnetic anisotropy, similar as in \cite{BEV19}. Two symmetric tilt grain boundaries are constructed in a periodic domain, see Figure \ref{fig:gbVel}. This tilted grain boundary has an angle of approximately 0.2048 $\pi$. A magnetization field is chosen in the easy direction, \hkl<11>, of the grains to the left and right. As a result, the grain in the center has a larger bulk energy due to magnetic anisotropy, and the magnetic driving force on the grain boundaries points towards the center. As the grain boundaries move at a constant velocity, the free energy decays linearly. The steps in the free energy decay are a result of the atomistic configuration of the grain boundary. Ultimately, the grain boundaries annihilate and the free energy drop corresponds to the grain boundary energy in the system. As a result of the crystal's square symmetry, multiple equivalent directions of magnetization cause the same magnetic driving force. 

In model \ME, the elongation of the peaks indicates the direction of magnetization. We consider two cases. In one case, the magnetization is more parallel to the grain boundary (open diamond). In the other case, the magnetization is more perpendicular to the grain boundary (closed diamond). Despite the same magnetic driving force, the energy decay in both cases differ. This results in a difference in the velocity of the grain boundary. In the case of magnetizations that are more perpendicular to the grain boundary, the grain boundary moves faster. Consequently, the mobility, $M_f$, depends on the orientation of the magnetization relative to the grain boundary. Microscopically, this is due to the anisotropic peaks in the rescaled density field, as the structure of the grain boundary is clearly influenced by the orientation of the elongated peaks.
In spite of this, the final step in the decay of energy is only slightly affected by the direction of magnetization, and the energy at the grain boundary is nearly the same regardless of the direction of magnetization. 

There is no dependence between the orientation of magnetization and the free energy decay or rescaled density distribution in \MXfour (open and closed squares). There is no difference in the velocity of the grain boundary between the two scenarios. This means that the mobility of the grain boundary is independent of the orientation of the magnetization. It should also be noted that the grain boundary energy is similar to that of model \ME.

\begin{figure}[htb]
  \medskip
  \begin{tabular}{cc}
    \includegraphics*[width= 0.24 \textwidth]{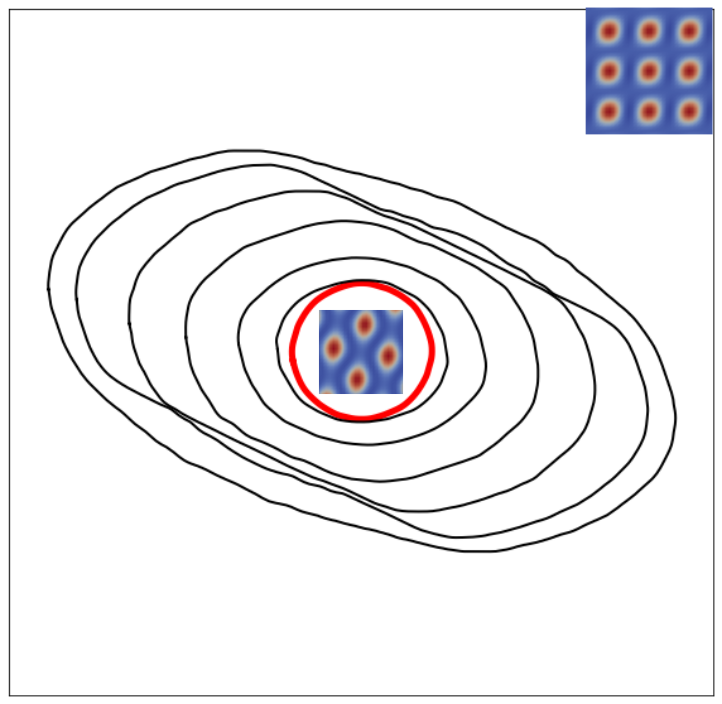} &
\includegraphics*[width= 0.24 \textwidth]{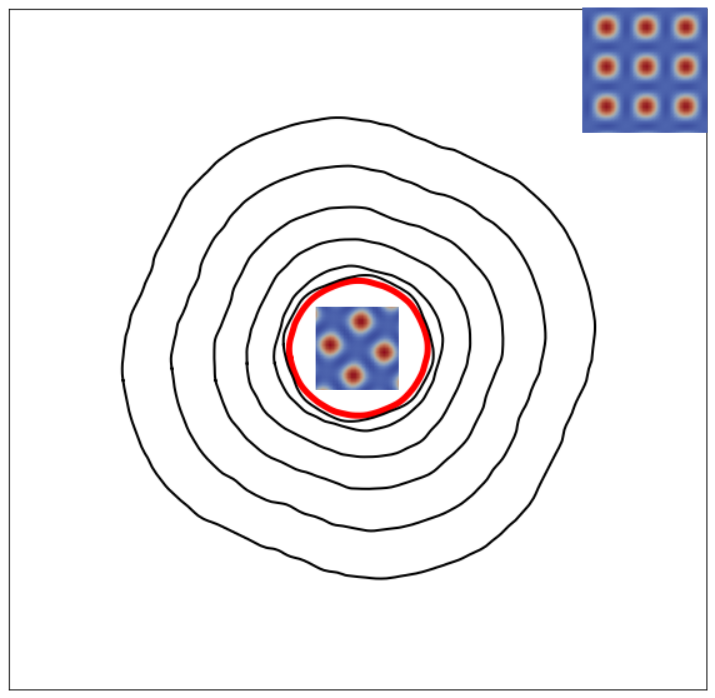} 
  \end{tabular}
  \medskip
  \caption{ \label{fig:shape} Magnetically driven grain growth for model \ME (left) and \MXfour (right). A circular grain is constructed (red isoline). The magnetization corresponds to the easy direction of the grain which points upward. The inlays show the structure in the grain and in the matrix. } 
  \end{figure}

Figure~\ref{fig:shape} shows the magnetically driven growth of a circular grain. Initially, a circular grain with a radius of 40 is constructed. Compared to the outer matrix, the grain is rotated by 0.2083 $\pi$. A zoom of the system's rescaled density field can be seen in the inlays. The circular grain is magnetized in the \hkl[11] direction. Consequently, magnetic anisotropy exerts an outward driving force on grain boundaries. The magnetic driving force is large enough to overcome the force resulting from mean curvature flow in this case and the grain begins to grow. The grain elongates perpendicular to the magnetization in model \ME. The anisotropic mobility discussed above accounts for this. The grain grows symmetrically in model \MXfour. There is a change in the shape of the initially spherical grain to a square grain with rounded corners. In this case, the sides of the square correspond to the boundaries between the grain and the matrix that are symmetrically tilted. There is no break in the 4-fold symmetry of the crystal due to magnetization.

The 2-fold symmetry of the grain for \ME can be attributed to the anisotropic mobility of the grain, since the grain boundary energy does not have this anisotropy, see Figure~\ref{fig:gbVel}. However, for X4 the anisotropic growth shapes can result from the anisotropy of the grain boundary energy or an anisotropic mobility. 

\subsection{Thin film coarsening}

\begin{figure*}[htb]
\medskip
 \includegraphics*[width= 0.95 \textwidth]{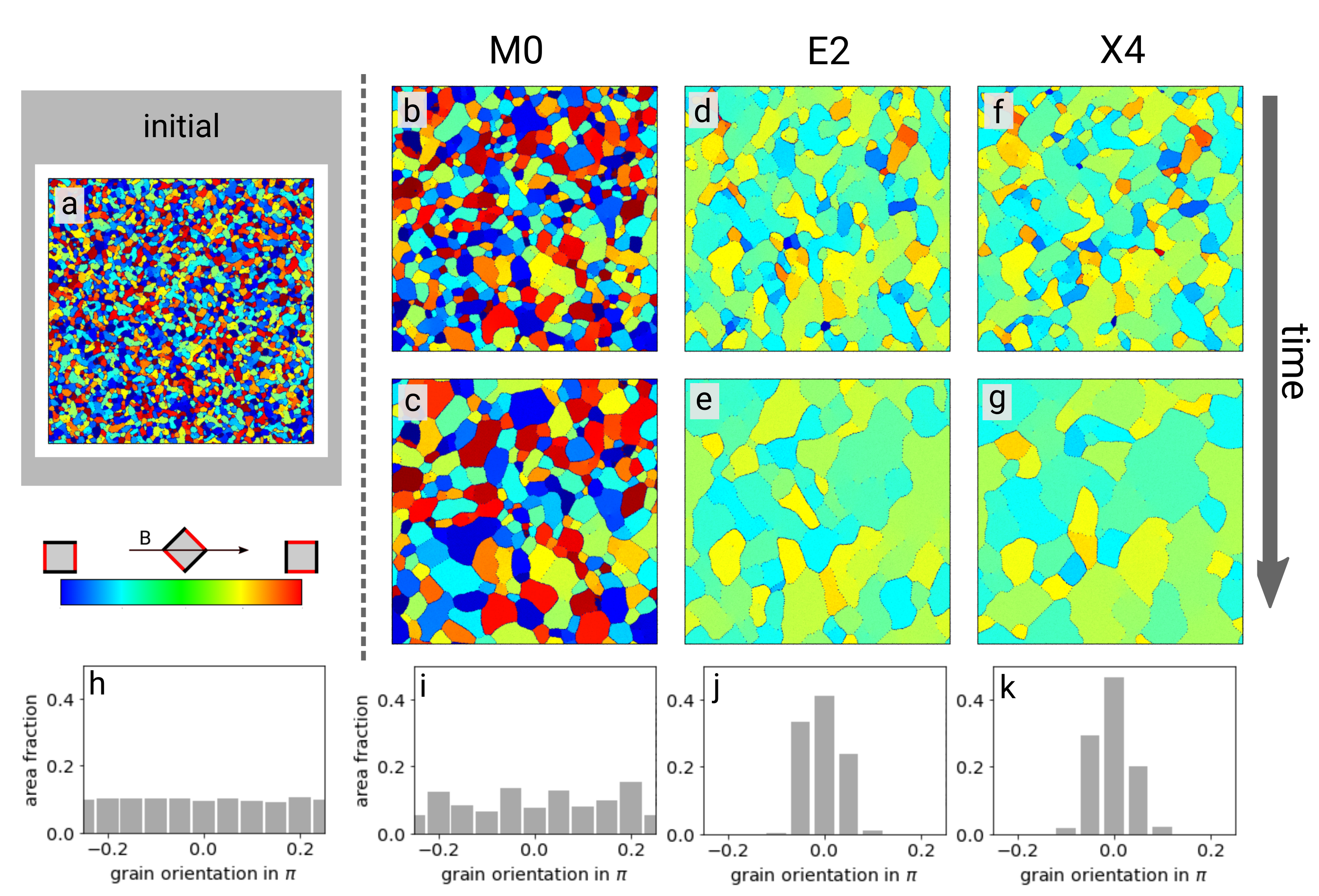}
  \medskip
  \caption{\label{fig:orient}(a) The initial configuration of the coarsening simulation. The color indicates the crystal's orientation in relation to the external magnetic field. The external magnetic field is oriented in the x direction. Grains with \hkl<11> in magnetization direction are green. The initial orientation distribution is isotropic (h). 
(b-g) Simulation of coarsening based on different setups and two different time snapshots. (i-k) Orientation distribution after coarsening. "ref" corresponds to no magnetization. \ME and \MXfour correspond to parameter set (a) in Table~\ref{tab:mpar}. As a result of magnetization, grain selection occurs for \ME and \MXfour (j,k). The computational domain is 819 × 819.  
}
\end{figure*}

This section examines the impact of different magnetic coupling approaches on thin film coarsening. Initially, small grains are randomly distributed, as shown in Figure~\ref{fig:orient}. Over time larger grains grow at the expense of smaller grains and coarsening occurs. 

We are first interested in the orientation of the grains. The orientation distribution remains isotropic without magnetization, which means that coarsened grains have no preferred orientation, see Figure ~\ref{fig:orient}(b,c,i). With magnetization this changes, see Figure ~\ref{fig:orient}(d,e,j) and (f,g,k). This phenomenon has already been described in \cite{BEV19,BV20}. The observed grain selection is a result of the magnetic anisotropy, which creates an additional driving force that causes grains oriented parallel to the magnetization, which have the lowest energy, to grow at the expense of grains with different orientations. As a result the orientation distribution is dominated by grains oriented parallel to the magnetization. In Figure~\ref{fig:orient} grains with a preferred alignment are shown in green and dominate the final texture. As \ME and \MXfour have the same magnetic anisotropy, their orientation distributions are similar.

\begin{figure}[htb]
\medskip
\includegraphics*[width = 0.45 \textwidth]{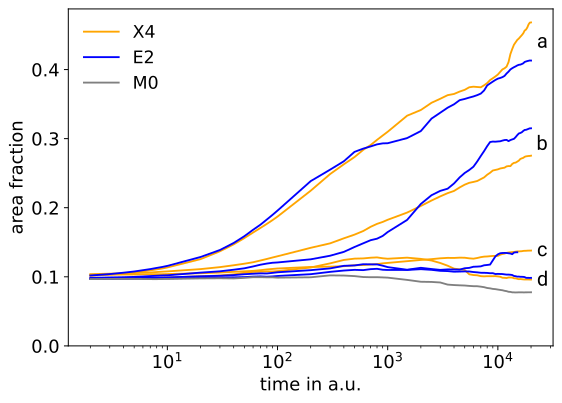}
  \medskip
  \caption{\label{fig:select} Grain selection. 
The area fraction of grains in which the direction of one \hkl[11] crystal plane does not deviate more than $\pm 0.025 \pi$ from the direction of magnetization. For a perfectly isotropic orientation distribution, the area fraction is 0.1. Labels (a)-(d) correspond to the model parameters, see Table~\ref{tab:mpar}. The results are shown in comparison with no magnetic driving froce (M0).
}
\end{figure}

The evolution over time of this grain selection process is shown in Figure~\ref{fig:select}, which illustrates the area fraction of grains in which one of the \hkl[11] directions does not deviate more than $\pm 0.025 \pi$ from the magnetization direction. In the absence of magnetization, orientation distributions are always isotropic. Therefore, the area fraction does not change significantly and is around 0.1, see Figure~\ref{fig:orient}(i). The area fraction is only minimally affected by a small magnetic driving force, as for parameters (d) and (c) in Table \ref{tab:mpar}. For larger driving forces, as for parameters (b) and (a), the area fraction grows, with a growth-rate proportional to the magnetic driving force. Not only the final distribution is similar, Figure~\ref{fig:orient}(j,k) for case (a), also for the evolution E2 and X4 lead to similar grain selection for all considered magnetic driving forces. 

\begin{figure}[htb]
\medskip
\includegraphics*[width = 0.45 \textwidth]{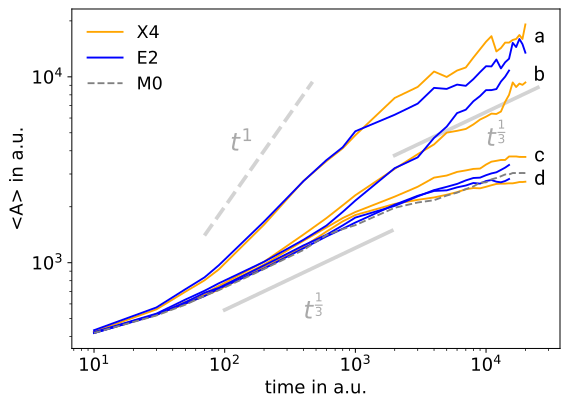}
  \medskip
  \caption{\label{fig:scaling} Scaling of mean grain size $\langle A \rangle $ for parameters (a) - (d) in Table \ref{tab:mpar} together with M0 corresponding to no magnetic driving force. Different coarsening regimes are identified: dependent scaling, a magnetically enhanced
scaling regime with the scaling exponent depending on the magnetic driving force varying between $1/3$ and $1$; independent scaling, a regime reached at late
times, with a scaling exponent of $1/3$ independent of magnetic anisotropy for large magnetic driving (parameters (a) and (b)); towards stagnation, a regime which is only present without or with low
magnetic driving (parameters (c) and (d)).}
\end{figure}

Next we are concerned with possible scaling of the coarsening process. The complex influence of a magnetic field has already been explained in \cite{BV20}. We essentially observe the same behaviour with both models E2 and X4. The coarsening is enhanced by increasing the magnetization. In all cases we identify different scaling
regimes for the mean grain size $\langle A \rangle \propto t^\beta$, with a scaling exponent $\beta$. After an initial phase a regime is reached for which $\beta$ depends on the magnetic driving force. Without magnetization, referred as M0 in Figure \ref{fig:scaling}, $\beta = 1/3$ is found. Also for low magnetic driving, corresponding to parameters (d) and (c) in Table \ref{tab:mpar}, this exponent is not significantly affected. For larger driving forces, corresponding to parameters (b) and (a) in Table \ref{tab:mpar}, the exponent increases, approaching $\beta = 1$. The additional driving force, if large enough,  overcomes the intrinsically present retarding forces, such as triple point and orientational pinning, and thus leads to an enhanced coarsening, with an exponent depending on the magnetic driving force. However, this scaling regime ends. For small magnetic driving, parameters (d) and (c), it turns into stagnation. For large magnetic driving, parameters (b) and (a), coarsening proceeds but with a constant, independent of the magnetic driving forces, exponent $\beta = 1/3$. This behaviour can be explained by the described grain selection. It is this grain selection which decreases the mean magnetic driving force over time. If the texture is dominated by well aligned grains, the magnetic driving force is limited by the texture and only parts of the retarding forces can be overcome. Low magnetic driving forces do not overcome the retarding force, which leads to stagnation. Also the scaling behaviour, with the different regimes and the exponents, are similar for E2 and X4.

\begin{figure}[htb]
\medskip
\includegraphics*[width = 0.45 \textwidth]{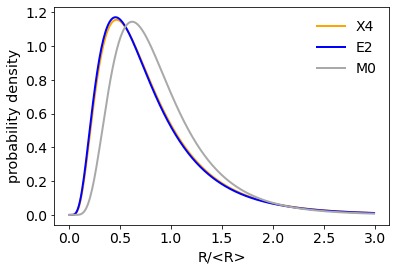}
  \medskip
  \caption{\label{fig:GSD} Grain size distribution (GSD). A large magnetic driving force corresponding to parameters (a) in Table \ref{tab:mpar} is considered in comparison with no magnetization (M0).
}
\end{figure}

\begin{figure*}
  \begin{tabular}{ccc}
    \raisebox{-.5\height}{ a)\includegraphics[width=0.3\textwidth]{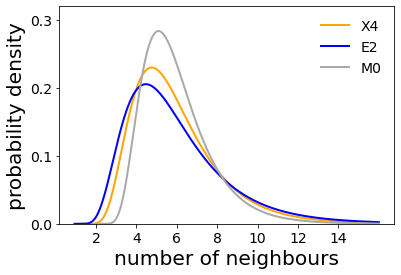}} &
    \raisebox{-.5\height}{ b)\includegraphics[width=0.3\textwidth]{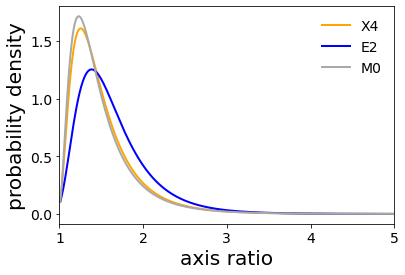}} &
    \raisebox{-.5\height}{ c)\includegraphics[width=0.3\textwidth]{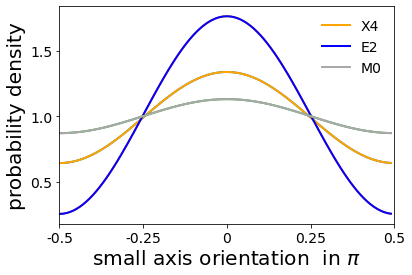}}
\end{tabular}
\caption{\label{fig:XXD} Grain coordination and shape. a) Log-normal description for next neighbor distribution (NND), the smoothed distribution should of course be interpreted in a discrete setting. b) Log-normal description for axis ratio distribution (ARD). c) Cosine description for small axis orientation distribution (SAOD). All data are obtained from late time coarsening regime for parameter (a) in Table \ref{tab:mpar} in comparison with no magnetization (M0). Individual data and further parameters are provided in Figure \ref{fig:sup}.}
\end{figure*}

Until now the different modeling approaches to incorporate the magnetic driving forces in models E2 and X4 did not lead to any significant change in thin film coarsening. 
Next we examine various geometrical and topological properties of the grain structure. This is done within the independent scaling regime where self similar growth
is observed. We first consider the grain size distribution (GSD), see Figure~\ref{fig:GSD}.
Magnetically enhanced coarsening shifts the peak of the log normal distribution to a smaller value. Thus, the number of large grains relative to the average grain size is increased. As already discussed in \cite{BV20} this behaviour corresponds with experimental results on Zr sheets \cite{MB10}. The behaviour is the same for E2 and X4. Next we examine the next neighbor distribution (NND), the axis ratio distribution (ARD) and the small axis orientation distribution (SAOD), see Figure \ref{fig:XXD}.
NND or coordination number of grains counts the number of neighboring grains. ARD considers the shape of the grains, which is quantified by approximating every grain by an ellipse and considering the ratio of the axis of the ellipse as a measure of elongation. To consider the direction of elongation we measure the angle of the small axis with the external magnetic field. This defines SAOD.
NND and ARD are best fitted by a log-normal distribution and SAOD by a cosine function. For NND the magnetic driving force leads to a broader distribution and the maximum is shifted to smaller values. This can be related to the faster growth, which leads to larger grains and thus also an increased difference in grain size. Classical empirical laws for topological properties in grain structures, postulate a linear relation between the coordination number and the area of the grains. These effect is further enhanced by the elongation of grains, which lead to more neighbors and explains the broadening. The broadening is less significant for X4. ARD significantly changes only for E2, for which the ratio
increases, which shows that more elongated grains are present. In contrast to this the magnetic driving force in X4 does not significantly alter ARD, the shape of the grains is similar to the once in the late stage of the coarsening process without magnetic driving force. This difference also explains the difference between E2 and X4 in NND. As a change in grain shape due to magnetization has not been observed experimentally and is also hard to explain theoretically this difference between E2 and X4 is an important advantage of X4. This is further enhanced by considering SAOD, which measures the correlation of the elongation with the external magnetic field. While the elongated grains are strongly oriented perpendicular to the external magnetic field in E2, this effect is much less pronounced for X4. 



\section{Summary}

We studied magnetically enhanced coarsening of thin films using a magnetic XPFC model. The external magnetic field is assumed to be strong enough for the magnetization to be constant and perfectly aligned with the external magnetic field. We account for the magneto-structural effects by modifying the correlation function in reciprocal space and compare two modeling approach. The first builds on \cite{BSV22b} and adds to the usual correlation function $\Ch_{\rm XPFC}$ the lowest order expansion of the coupling terms between the rescaled density $\phii$ and the magnetization $\mm$. This approach leads to the same form as considered in \cite{BV20}. The second directly modifies the Gaussians used in defining $\Ch_{\rm XPFC}$ by considering an angle dependent height in relation to the magnetization. The approaches are termed E2 and X4, respectively. They are parameterized to lead to the same magnetic anisotropy. We compare both models for various situations. The anisotropy of the magnetic properties of the crystal lead to a magnetic driving force. Well aligned crystals grow at the expense of not well aligned crystals. This leads to an enhanced coarsening if the magnetic driving force is large enough. In addition we observe grain selection, which leads to texture dominated by well aligned grains. As this reduces the mean orientation difference between grains the mean magnetic driving force decreases and leads to a turn over in the coarsening regime, from a dependent scaling with $\langle A \rangle \sim t^\beta$ with $\beta$ between $1/3$ and $1$ to an independent scaling with $\beta = 1/3$ or stagnation, depending on the strength of the applied magnetic field. While these properties and also the shift in grain size distribution (GSD) due to the magnetic driving force are the same for both models and in agreement with experimental data, other geometrical and topological properties differ. Most significantly the aspect ratio distribution (ARD) and the small angle orientation distribution (SAOD) differ. While E2 shows a significant elongation of grains in the direction perpendicular to the magnetic field, this is not the case for X4. This difference also has an effect on the next neighbor distribution (NND) and can be explained by microscopic differences. Detailed investigation of planer grain boundaries and circular grains demonstrate an orientation dependent mobility for E2, which can be attributed to anisotropically deformed peaks in the rescaled density field, which is the main source for grain elongation. Also the magnetostriction effect leads to deformation of crystal and defect structures. Within E2 magnetic anisotropy, anisotropic grain boundary mobility and magnetostriction cannot be decoupled. Model X4 accounts for magnetic anisotropy without magnetostriction. The peaks in the rescaled density field are not deformed. As a result it reproduces the main properties of thin film coarsening. Not only the scaling regime and the GSD are in agreement with experimental results, but also all other considered geometrical and topological properties, NND, ARD and SAOD, show the envisioned influence of the magnetic field. Combining both models would in principle allow to control magnetic anisotropy and magnetostriction independently, to reproduce magnetostructural properties of specific materials.


\section{Appendix}

\begin{figure*}[htb]
  \begin{tabular}{ccc||c}
    & E2 & X4 & M0 \\
   a & \includegraphics[width= 0.3 \textwidth]{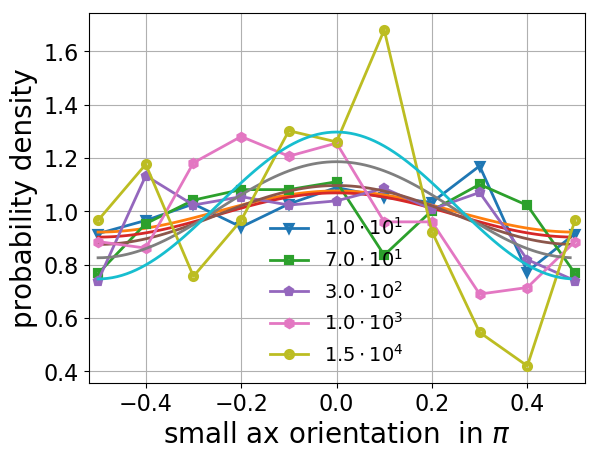} 
     & \includegraphics[width= 0.3 \textwidth]{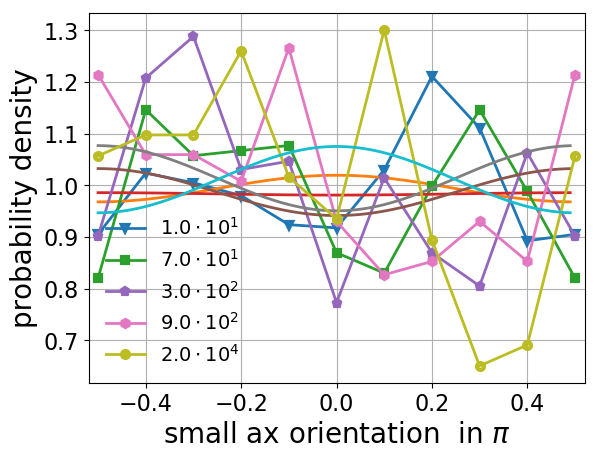}
      & \\
    b & \includegraphics[width= 0.3 \textwidth]{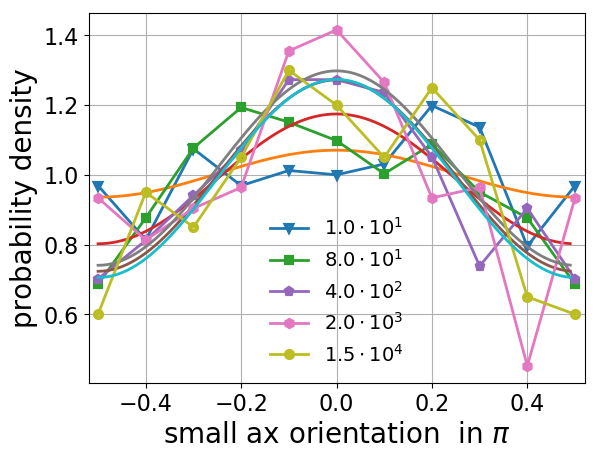}
      & \includegraphics[width= 0.3 \textwidth]{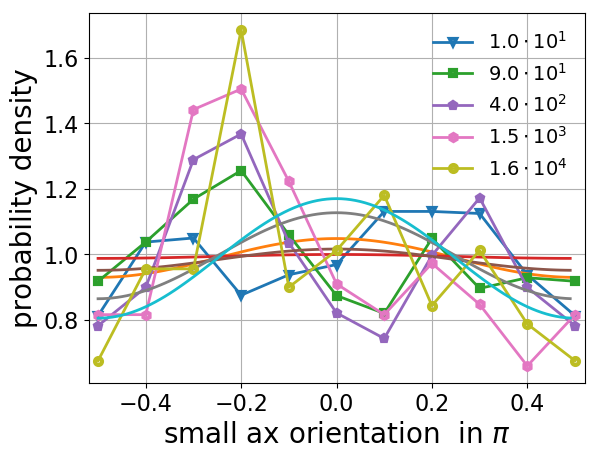}
      & \\
    c & \includegraphics[width= 0.3 \textwidth]{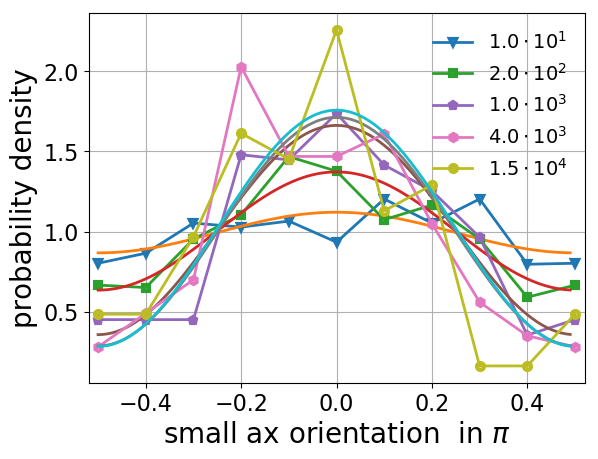}
      & \includegraphics[width= 0.3 \textwidth]{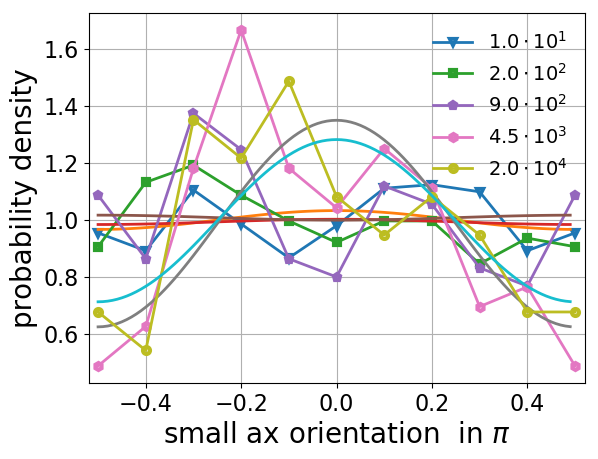}
      & \\
    d & \includegraphics[width= 0.3 \textwidth]{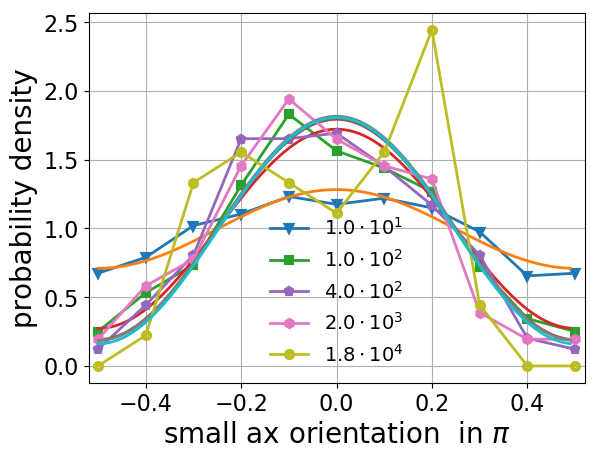}
      & \includegraphics[width= 0.3 \textwidth]{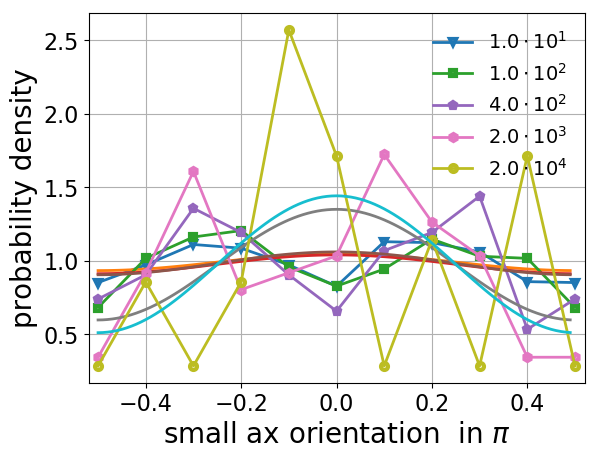}
      & \includegraphics[width= 0.3 \textwidth]{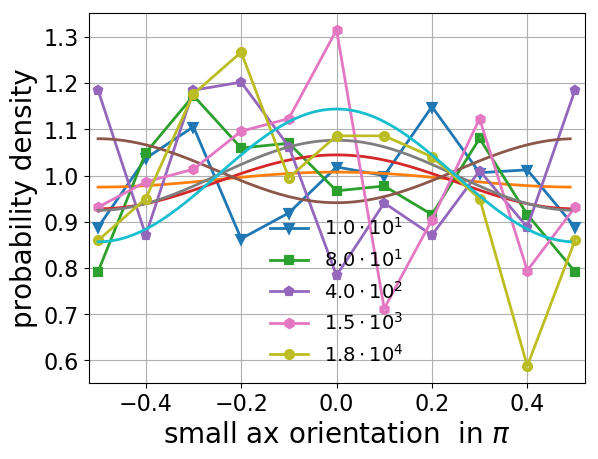}\\
  \end{tabular}
  \caption{ The raw data of small axis orientation (SAOD) is displayed, showing the probability of finding a grain with its small axis aligned with the magnetization for different magnetic anisotropy, accoring to parameters (a)-(d) in Table \ref{tab:mpar}, see fig.~\ref{fig:MAniso}. Every plot shows the distribution at different times during coarsening. In model E2, the orientation distribution of the small axis becomes more concentrated towards alignment with the magnetization as time and magnetic anisotropy increase. There is no apparent trend in model X4. As expected, without magnetization (M0) no effect is visible.
  \label{fig:sup} }
\end{figure*}

We provide individual data from which the small axis orientational distribution (SAOD) in Figure \ref{fig:XXD} is obtained. They result from the late time coarsening regime. Figure \ref{fig:sup} show the raw data over time for all parameters (a) - (d) in Table \ref{tab:mpar}. The distribution in Figure \ref{fig:XXD} is obtained by averaging over the time interval $[5\cdot10^3,2\cdot10^4]$. 

\medskip

\medskip
\textbf{Acknowledgements} \par 
AV and RB acknowledge support by the German Research Foundation (DFG) within SPP1959 under Grant No. VO899/20-2. We further acknowledge computing resources provided at J\"ulich Supercomputing Center under grant PFAMDIS.

\medskip

%
\bibliographystyle{MSP}
\bibliography{allDiss}


\begin{figure}
\textbf{Table of Contents}\\
\medskip
  \includegraphics{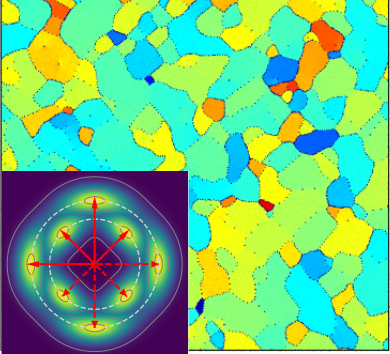}
  \medskip
  \caption{Thin film coarsening under the influence of a strong magnetic field. Large scale extended phase field crystal (XPFC) simulations are considered which account for magneto-structural interactions by a magnetically modified correlation function shown in reciprocal space for a square lattice.}
\end{figure}


\end{document}